\documentclass{ptephy}
\preprintnumber{XXXX-XXXX}

\usepackage{boites}
\numberwithin{equation}{section}
\usepackage{amsmath}	

\usepackage{amsmath}
\usepackage{amssymb}

\makeatletter
\DeclareRobustCommand{\cev}[1]{%
  \mathpalette\do@cev{#1}%
}
\newcommand{\do@cev}[2]{%
  \fix@cev{#1}{+}%
  \reflectbox{$\m@th#1\vec{\reflectbox{$\fix@cev{#1}{-}\m@th#1#2\fix@cev{#1}{+}$}}$}%
  \fix@cev{#1}{-}%
}
\newcommand{\fix@cev}[2]{%
  \ifx#1\displaystyle
    \mkern#23mu
  \else
    \ifx#1\textstyle
      \mkern#23mu
    \else
      \ifx#1\scriptstyle
        \mkern#22mu
      \else
        \mkern#22mu
      \fi
    \fi
  \fi
}

\usepackage{version}	
\begin{document}
\title{
Relation of 
$a^\dagger a$ terms to higher-order terms 
in the adiabatic expansion
for large-amplitude collective motion 
}

\author{\name{\fname{Koichi} \surname{Sato}}{1} 
}

\address{\affil{1}{Department of Physics, Osaka City University, Osaka 558-8585, Japan}
\email{satok@sci.osaka-cu.ac.jp}}

\begin{abstract}
We investigate the relation of $a^\dagger a$ terms in the collective
operator to the higher-order terms in the adiabatic self-consistent
collective coordinate (ASCC) method.
In the ASCC method, a state vector is written as $e^{i\hat G(q,p,n)}|\phi(q)\rangle$ 
with $\hat G(q,p,n)$ which is a function of collective coordinate $q$, its conjugate momentum $p$
and the particle number $n$.
According to the generalized Thouless theorem, $\hat G$ can be written as a linear combination of two-quasiparticle creation and
annihilation operators $a^\dagger_\mu  a^\dagger_\nu$ and $a_\nu a_\mu$.
We show that, if $a^\dagger a$ terms are included in $\hat G(q,p,n)$,
it corresponds to the higher-order terms in the adiabatic expansion of $\hat G$.
This relation serves as a prescription to determine the higher-order collective operators
from the $a^\dagger a$ part of the collective operator, once it is given
without solving the higher-order equations of motion.
\end{abstract}

\subjectindex{xxxx, xxx}

\maketitle

\section{Introduction}
According to the generalized Thouless theorem
(Refs. \cite{Thouless1960,Marumori1980,Rowe1980,
Ring1977, Suzuki1983}),
a Hartree--Fock--Bogoliubov-type state vector
(a generalized Slater determinant) can be written in a unitary form as
\begin{align}
 |\psi\rangle =e^{i\hat G}|\phi\rangle,\,\,
\hat G=\sum_{\mu\nu} \left(
Z_{\mu\nu}a^\dagger_\mu a^\dagger_\nu + Z_{\mu\nu}^*a_\nu a_\mu\right),
\end{align}
where $|\phi\rangle$ is the vacuum with respect to quasiparticle
operator ${a_\mu}$ ($a_\mu|\phi\rangle =0$).
The purpose of this paper is to investigate the role of $a^\dagger a$
terms, if included in $\hat G$, in the context of adiabatic
approximation to time-dependent Hartree--Fock--Bogoliubov (TDHFB) theory.
In a recent paper (Ref. \cite{Sato2017}),
we analyzed the higher-order collective coordinate operators 
and their roles in the  gauge invariance
of the adiabatic self-consistent collective coordinate~(ASCC) method
(Ref. \cite{Matsuo2000}), 
which can be regarded as an advanced version of the adiabatic TDHFB~(ATDHFB) theory.

In this paper, we investigate the relation between the $a^\dagger a$ terms 
and the higher-order collective coordinate operators in the adiabatic expansion. 
We shall call $a^\dagger a^\dagger$ and $aa$ terms A-terms and 
$a^\dagger a$ and $aa^\dagger$ (or equivalently $a^\dagger a$ and constant terms)
B-terms, respectively.
With this terminology, 
the generalized Thouless theorem states that
$\hat G$ is uniquely given by a linear combination of A-terms.
The (generalized) Thouless theorem is useful to 
express the Hartree--Fock(--Bogoliubov)-type state vectors and 
plays an important role in the time-dependent mean-field theory (Ref. \cite{Ring1980}).

In contrast with the theorem, in the 
ASCC method, B-terms were introduced in $\hat G$ (Refs. \cite{Hinohara2007, Hinohara2008,Hinohara2009}), 
and it is closely related to the gauge invariance of the theory as we shall explain below.
In Ref. \cite{Sato2017}, we analyzed the gauge symmetry and
its breaking in the ASCC method, and showed that 
the gauge invariance is (partially) broken by the adiabatic approximation
and that one needs 
the higher-order collective operators to retain the gauge invariance.
In this paper, we show that the introduction of B-terms in $\hat G$ is equivalent to
that of a certain kind of higher-order operators, which are written
in terms of the B-terms. 

The ASCC method (Ref. \cite{Matsuo2000}) is a practical method for describing the large-amplitude 
collective motion of atomic nuclei with superfluidity.
It is an adiabatic approximation to the self-consistent collective
coordinate (SCC) method, which
was originally formulated by Marumori et al~(Ref. \cite{Marumori1980}),
and can be regarded as an advanced version of the ATDHFB theory.
The ASCC method overcomes the difficulties, 
which several versions of the ATDHF(B) theory encountered [see
Refs. \cite{Klein1991,Matsuyanagi2010, Nakatsukasa2016} for a review],
and enables one to describe the large-amplitude collective dynamics which cannot be treated by
the $(\eta,\eta^*)$ expansion of the SCC method.

The gauge symmetry in the ASCC method was first pointed out by Hinohara
et al~(Ref. \cite{Hinohara2007}).
They encountered numerical instability in the calculation of the
one-dimensional ASCC method. (We mean by the $D$-dimensional ASCC method
that the dimension of the collective coordinate $q$ is $D$.) 
They found that the instability  was caused by
the symmetry associated with some continuous transformation under which the basic equations
of the theory are invariant. As the transformation changes the phase of the state vector,
they called this symmetry the ``gauge'' symmetry and
proposed a prescription for the numerical stability (``gauge fixing''),
which led them to successful calculation.
After the successful application of the one-dimensional ASCC method by Hinohara et al.,
an approximate version of the two-dimensional ASCC method, which is called
the constrained Hartree--Fock--Bogoliubov (HFB) plus local quasiparticle random phase approximation (QRPA) method, was developed and
applied to a variety of quadrupole shape dynamics~
(Refs.
\cite{Sato2011,Watanabe2011,Hinohara2011a,Hinohara2011,Hinohara2012,Yoshida2011,
Sato2012}).
However, little progress had been made in the understanding of the gauge symmetry.
Recently, we analyzed the gauge symmetry in the ASCC method 
on the basis of the Dirac--Bergmann theory of the constrained systems
(Refs. \cite{Dirac1950,Anderson1951,Dirac1964}),
which shed a new light (Refs. \cite{Sato2015, Sato2017}).
It is worth mentioning that the one-dimensional ASCC method without pairing correlation was 
also successfully applied to the nuclear
reaction studies (Ref. \cite{Wen2016}).

In the ASCC method, a state vector is written in the form of
\begin{align}
|\phi(q,p,\varphi,n)\rangle = e^{-i\varphi \hat N} e^{i\hat G(q,p,n)}|\phi(q)\rangle.
\end{align}
Here,  $(q,p)$ are the collective coordinate and conjugate collective momentum.
$n$ is the particle number measured by the mean value $N_0$ ($n=N-N_0$),
and $\varphi$ is the gauge angle conjugate to $n$.
Hinohara et al~(Refs. \cite{Hinohara2007, Hinohara2008, Hinohara2009}) 
employed $\hat G$ expanded up to the first order of $(p,n)$: 
$\hat G=p\hat Q (q)+n \hat \Theta(q)$.
As mentioned above, they encountered the numerical instability, and
their prescription for the numerical stability is as follows.
They require the commutativity of the collective-coordinate and the
particle-number operators $[\hat Q,\hat N]=0$ for the gauge symmetry of  
moving-frame HFB \& QRPA equations, which are the equations of motion in
the ASCC method, and then fix the gauge.
However, the requirement of  $[\hat Q,\hat N]=0$ implies that 
they needed to include B-terms in $\hat Q(q)$, in contrast with
the original formulation of the ASCC method in Ref. \cite{Matsuo2000}, 
which respects the generalized Thouless theorem.

In Ref. \cite{Sato2017}, we showed that the gauge symmetry in the ASCC method
is broken by two sources: 
the decomposition of the equation of collective submanifold depending on
the order of $p$
and the truncation of the adiabatic expansion of $\hat G$ to a certain
order of $(p, n)$.
We showed that the gauge symmetry broken by the truncation is retained 
by including the higher-order operators as in Eq. (\ref{eq: G expanded up to 3rd}).
In this approach with the higher-order operators, the condition $[\hat Q,\hat N]=0$
is not necessary for the gauge symmetry, 
and one does not need B-terms in the collective operators .

Thus, there are two approaches to conserve the gauge symmetry.
One is the approach with higher-order operators 
consisting of only A-terms, and the other with only the first-order operator containing B-terms
as well as the A-terms, requiring $[\hat Q,\hat N]=0$.
Let us call the former Approach A and the latter Approach B.
Note that, as shown in Ref.  \cite{Sato2017},
the gauge symmetry in the canonical-variable conditions, which are
conditions
for the collective variables to be canonical,  is broken in Approach B,
while it is not in Approach A.
It is noteworthy that, in Approach A, the collective operators consists of A-terms
but that the gauge transformation mixes A-terms and B-terms.

When the ASCC method is applied to the translational motion,
the collective coordinate and momentum operators
$\hat Q$ and $\hat P$  correspond to the center-of-mass
position and momentum, respectively, and their exact operator forms are known.
Whereas the state vector can be written without the B-terms according
to the generalized Thouless theorem,
$\hat Q$ for the translational motion contains B-terms in the quasiparticle representation. 
With the B-terms included, $\hat G$ expanded up to the first order 
would give the exact solution for the translational motion.
Although the state vector could be written without B-terms,
if the B-terms are neglected, one might need to take into account 
the higher-order operators at the level of the equations of motion
after the adiabatic expansion.
Thus, it is no trivial whether or not one should 
include B-terms and/or higher-order terms in $\hat G$.
To address this point, one must investigate the relation between
the two approaches.

Concerning the higher-order terms and B-terms in the adiabatic
expansion for large-amplitude collective motion, 
the following two things are worthy of note.
First, in his paper on the ATDHF in 1977 (Ref. \cite{Villars1977}),
Villars mentioned the extension of his ATDHF theory 
including the higher-order operator 
(more strictly, 
the extension with the first- and third-order operators and no second-order operator)
and preannounced a publication on it
: 
''Ref. 17) A. Toukan and F. Villars, to be published'' in Ref. \cite{Villars1977}. 
However, as far as the author knows, it was not published after all.
Second, in the ATDHF theory by Baranger and V\'en\'eroni (Ref. \cite{Baranger1978}),
they proposed the density matrix in the form of
$\rho(t)=e^{i\chi(t)}\rho_0 (t) e^{-\chi(t)}$ with Hermitian and time-even 
$\rho_0(t)$ and $\chi(t)$.
They emphasized that $\chi(t)$ can be written in terms of A-terms only,
but included B-terms as well as A-terms in the treatment 
of the translational motion.
In this paper, we attempt to clarify the relation between 
the higher-order terms and B-terms.

The paper is organized as follows.
In Sect. 2, after giving a brief explanation of the formulation of the
ASCC method, we compare the moving-frame HFB \& QRPA equations
between the two approaches. We shall find some correspondence
between the higher-order operators and the (multiple) commutators 
of the A-part and B-part of the first-order operators.
%
%
This comparison  is useful for understanding  
the contribution of the B-part of the collective coordinate operator
$\hat Q(q)$ to the equations of motion.
In Sect. 3, we illustrates how to obtain the correspondence between the higher-order operators
and the multiple commutators of the first-order operators in general.
By comparing the state vectors in the two approaches directly,
we show that the inclusion of B-terms is equivalent to that of 
a certain kind of the higher-order operators
and give the explicit expression of the corresponding higher-order operators.
This correspondence gives a prescription to determine the higher-order collective 
operators from the B-part of the first-order collective coordinate operator $\hat Q_B(q)$.
In Sect.4, we compare the inertial masses and confirm that,
if we determine the higher-order operators with the above-mentioned prescription,
the two approaches give the same results. 
Concluding remarks are given in Sect. 5.
In Appendix, some formulae of commutators of fermion operators are
given, which helps understand the derivation in the text.

\section{Comparison of the equations of motion}
In this section,  we give a minimal explanation of the formulation
of the ASCC method which is necessary for the purpose of  this paper.
(For details, see Refs. \cite{Matsuo2000, Sato2017}.)
Then we shall compare the moving-frame HFB \& QRPA equations
in the two approaches. 
Although we compare the state vectors in the two approaches directly in
the next section,
the comparison of the moving-frame equations in this section is useful 
to understand  how the B-part of $\hat G$
contribute to the equations of motion.

The state vector in the ASCC method is written as
\begin{align}
|\phi(q,p,\varphi,n)\rangle = e^{-i\varphi \hat N} e^{i\hat G(q,p,n)}|\phi(q)\rangle .
\label{eq:state vector}
\end{align}
We assume the $\varphi$-dependence of the state vector as above, which
guarantees the conservation of the expectation value of the particle
number $\hat N$.
Although there are two components, neutrons and protons, in atomic nuclei, 
we consider the ASCC method with a single component for simplicity.
We show below $\hat G$ expanded up to the third order in Approach A, 
\begin{align}
\hat G(q,p,n) &=p \hat Q^{(1)}(q)+n \hat \Theta^{(1)}(q)
+\frac{1}{2}p^2 \hat Q^{(2)}(q) 
+\frac{1}{2}n^2 \hat \Theta^{(2)} (q) +pn \hat X \notag\\
&+\frac{1}{3!}p^3 \hat Q^{(3)}(q) 
+\frac{1}{3!}n^3 \hat \Theta^{(3)}(q) 
+\frac{1}{2}p^2n \hat O^{(2,1)}(q) 
+\frac{1}{2}pn^2 \hat O^{(1,2)}(q) .
\label{eq: G expanded up to 3rd}
\end{align}
All the operators  in Eq. (\ref{eq: G expanded up to 3rd})
consist of A-terms only. 
For example, $\hat Q^{(i)}(q)$ is defined by
\begin{align}
\hat Q^{(i)}(q)= \sum_{\alpha \beta}  Q^{(i)}_{\alpha \beta}(q) a_\alpha^\dagger a_\beta^\dagger + 
Q^{(i)*}_{\alpha \beta}(q) a_\beta a_\alpha  \,\,\,(i=1,2,3).
\end{align}
The other operators are defined similarly.

In Approach B, $\hat G$ is expanded up to the first order as
\begin{align}
\hat G(q,p,n) &=p \hat Q(q)+n \hat \Theta(q) .
\label{eq: G expanded up to 1st}
\end{align}
We omit the superscripts indicating the order of expansion in Approach B.
In Approach B, while  $\hat \Theta(q)$ consists of only A-terms as in Approach A, 
$\hat Q$ contains B-terms.
\begin{align}
\hat Q(q) &=\hat Q_A(q)+\hat Q_B(q),    \\
\hat Q_A(q)&= \sum_{\alpha \beta}  Q_{A\alpha \beta}(q) a_\alpha^\dagger a_\beta^\dagger + 
Q^{*}_{A\alpha \beta}(q) a_\beta a_\alpha  ,\\
\hat Q_B(q)&= \sum_{\alpha \beta}  Q_{B\alpha \beta}(q) a_\alpha^\dagger a_\beta, \label{eq:Q_B}
\end{align}
where $Q_B$ is a Hermitian matrix.
[We denote the A(B)-part of an operator $\hat O$ by $\hat O_{A(B)}$ hereinafter.]
One might wonder 
if $\hat Q_B$ could be written in a more general form
\begin{align}
 \hat Q_B = \sum_{\mu\nu} Q_B^{\mu\nu}a_{\mu}^\dagger a_{\nu} +\sum_{\mu\nu} \tilde Q_B^{\mu\nu}a_{\mu} a_{\nu}^\dagger ,
\end{align}
with Hermitian matrices $Q_B$ and $\tilde Q_B$, and it can be rewritten as
\begin{align}
 \hat Q_ B& = 
 \sum_{\mu\nu} (Q_B^{\mu\nu} -\tilde Q_B^{\nu\mu } )a_{\mu}^\dagger a_{\nu} +
 \sum_\mu \tilde Q_B^{\mu\mu} .
\end{align}
$(Q_B -\tilde Q_B^{T} )$ is also Hermitian,
 and it implies that the right-hand side of Eq. (\ref{eq:Q_B})
could have a constant term.
However, the collective coordinate operator $\hat Q$ should satisfy 
the canonical-variable conditions (Ref. \cite{Matsuo2000}), and the zeroth-order
canonical-variable condition is given by
\begin{align}
\langle \phi (q) |\hat Q(q) | \phi(q)\rangle =\langle \phi (q) |\hat Q_B(q) | \phi(q)\rangle =0,
\end{align}
which implies that there is no constant term in Eq. (\ref{eq:Q_B}).

The equations of motion in the  ASCC method is derived from
the invariance principle of time-dependent Schr\"odinger equation
\begin{align}
 \delta\langle \phi(q,p,\varphi,n)|i\partial_t -\hat H 
|\phi(q,p,\varphi,n)\rangle = 0,
\label{eq:TDVP}
\end{align}
which is rewritten as the equation of collective submanifold,
\begin{align}
& \delta \langle \phi(q,p,\varphi, n)|\hat H 
- i\frac{\partial \mathcal{H}}{\partial p}\partial_q
- \frac{1}{i}\frac{\partial \mathcal{H}}{\partial q}\partial_p
- \frac{1}{i}\frac{\partial \mathcal{H}}{\partial \varphi}\partial_n
- \frac{\partial \mathcal{H}}{\partial n}\hat {N}
|\phi(q,p,\varphi, n) \rangle  &=0, \label{eq:Eq. of collective submanifold}
\end{align}
with the collective Hamiltonian $\mathcal{H}=\langle \phi(q,p,\varphi,n)|\hat H | \phi(q,p,\varphi,n)\rangle$.
We substitute the state vector (\ref{eq:state vector}), 
expand in powers of $(p,n)$ (adiabatic expansion), 
and decompose the above equation (\ref{eq:Eq. of collective submanifold}) depending on the order of $p$.
From the equations of $O(1)$,  $O(p)$ and $O(p^2)$,
the moving-frame HFB \& QRPA equations are derived, which are the equation
of motion in the ASCC method. 
%
%
When $\hat G$ is expanded up to $O(p^3)$ in Approach A,
the moving-frame HFB \& QRPA equations are given by

\noindent
\underline{Moving-frame HFB equation}
\begin{equation}
 \delta \langle \phi(q)|\hat H -\lambda \hat N -\partial_q V\hat Q^{(1)} |\phi(q)\rangle =0,
\label{eq:moving-frame HFB with Q3}
\end{equation}

\noindent
\underline{ Moving-frame QRPA equations}
\begin{equation}
 \delta \langle \phi(q)|[\hat H-\lambda \hat N -\partial_qV\hat Q^{(1)}, \hat Q^{(1)}] 
-\frac{1}{i}B(q)\hat P -\frac{1}{i}\partial_q V \hat Q^{(2)} |\phi(q)\rangle =0,
\label{eq:moving-frame QRPA1 with Q3}
\end{equation}

\begin{align}
 \delta \langle \phi(q)| &
[\hat H -\lambda \hat N -\partial_q V \hat Q^{(1)}, \frac{1}{i}\hat P]
-C(q)\hat Q^{(1)}-\partial_q \lambda \hat N \notag\\
&-\frac{1}{2B}
\partial_q V
\left\{[[\hat H-\lambda \hat N -\partial_qV\hat Q^{(1)},  \hat Q^{(1)}],\hat Q^{(1)}] -i [\hat H-\lambda \hat N, \hat
 Q^{(2)}] \right.\notag \\
&  \hspace{12em} 
\left. +\partial_q V( \hat Q^{(3)}-\frac{i}{2} [\hat Q^{(1)}, \hat Q^{(2)}])\right\}
|\phi(q)\rangle =0. \label{eq:moving-frame QRPA2 with Q3 }
\end{align}
Note that the moving-frame QRPA equation of $O(p)$ (\ref{eq:moving-frame
QRPA1 with Q3}) contains the second-order operator $\hat Q^{(2)}(q)$, 
and the moving-frame QRPA equation of $O(p^2)$ (\ref{eq:moving-frame
QRPA2 with Q3 }) does the third-order operator  $\hat Q^{(3)}(q)$ as
well as $\hat Q^{(2)}(q)$. 
Here $\hat Q^{(i)}\,(i=1,2,3)$ and $\hat P$ contain only A-terms.
Eqs. (\ref{eq:moving-frame HFB with Q3}) and (\ref{eq:moving-frame QRPA1
with Q3}) are derived from the $O(1)$ and $O(p)$
terms  of Eq. ($\ref{eq:TDVP}$), respectively.
Eq. (\ref{eq:moving-frame QRPA2 with Q3 }) are derived using 
the $O(1)$ and $O(p^2)$ terms. 

In Approach B, the moving-frame HFB \& QRPA equations are given by

\noindent
\underline{Moving-frame HFB equation}
\begin{equation}
 \delta \langle \phi(q)|\hat H -\lambda \hat N -\partial_q V\hat Q
  |\phi(q)\rangle =0,
\label{eq:moving-frame HFB}
\end{equation}

\noindent
\underline{ Moving-frame QRPA equations}
\begin{equation}
 \delta \langle \phi(q)|[\hat H-\lambda \hat N -\partial_q V \hat Q, \hat Q] 
-\frac{1}{i}B(q)\hat P  |\phi(q)\rangle =0,\label{eq:moving-frame QRPA1 }
\end{equation}

\begin{align}
 \delta \langle \phi(q)|&
[\hat H -\lambda \hat N -\partial_q V \hat Q, \frac{1}{i}\hat P]
-C(q)\hat Q-\partial_q \lambda \hat N \notag\\
&-\frac{1}{2B}
\partial_q V
\left\{[[\hat H-\lambda \hat N -\partial_q V \hat Q,  \hat Q],\hat Q] \right\}
|\phi(q)\rangle =0. \label{eq:moving-frame QRPA2 }
\end{align}
Note that $\hat Q$ is the first-order operator.

As shown in Ref. \cite{Sato2017}, Eqs.
(\ref{eq:moving-frame HFB with Q3})--(\ref{eq:moving-frame QRPA2 with Q3
}) are invariant under the following transformation:
\begin{align}
&\hat Q^{(1)} \rightarrow \hat Q^{(1)}     +\alpha \hat N, \\
&\hat Q^{(2)} \rightarrow \hat Q^{(2)}      +i\alpha [\hat N, \hat Q^{(1)}], \\
&\hat Q^{(3)} \rightarrow \hat Q^{(3)}     +\frac{3}{2}\alpha i [\hat N, \hat Q^{(2)}]      -\frac{1}{2}\alpha [\hat Q^{(1)},     [\hat Q^{(1)}, \hat N]], \\
& \lambda \rightarrow  \lambda -\alpha \partial_q V,\\
& \partial_q\lambda \rightarrow  \partial_q\lambda -\alpha C .
\end{align}

On the other hand, if $[\hat Q,\hat N]=0$,
Eqs. (\ref{eq:moving-frame HFB})--(\ref{eq:moving-frame QRPA2 }) are
invariant under the transformation 
\begin{align}
 \hat Q &\rightarrow \hat Q +\alpha \hat N , \\
 \lambda &\rightarrow  \lambda -\alpha \partial_q V,\\
 \partial_q\lambda &\rightarrow  \partial_q\lambda -\alpha C .
\end{align}
[Here, we have not shown the transformations of the operators which are not
involved in the moving-frame HFB \& QRPA equations
(\ref{eq:moving-frame HFB with Q3})--(\ref{eq:moving-frame QRPA2 })
above.
However, to consider the gauge symmetry in the canonical-variable
conditions, the transformations of the operators not shown here are needed.
See Ref. \cite{Sato2017} for the complete list of the transformations.
]

Before comparing the moving-frame equations between the two approaches, 
we shall give some remarks.
In the ASCC method, we take only the variation in the form of  
$\delta |\phi\rangle =a^\dagger_\mu a^\dagger_\nu|\phi\rangle$.
Therefore, the A-terms can directly contribute 
to the moving-frame HFB \& QRPA equation, 
but 
the variation of B-terms automatically vanishes.
\begin{align}
 \delta \langle \phi(q)|
\text{ (B-terms)}
|\phi(q)\rangle =0.  \label{eq:delta<B>=0}
\end{align}
The B-terms contribute only through commutators, e.g., $[\text{A-terms, B-terms}]$.
Concerning the commutators, the following rules are readily understood from
Eqs. (\ref{eq:[aa,a+a+]})--(\ref{eq:[a+a,a+a]}) and
are useful for the investigation below. 
\begin{align}
[\text{A-terms, A-terms}]&=\text{B-terms}, \label{eq:[A,A]=B}\\
[\text{A-terms, B-terms}]&=\text{A-terms},\\
[\text{B-terms, B-terms}]&=\text{B-terms}. \label{eq:[B,B]=B}
\end{align}
One can also see that the variation of the normally ordered forth-order
operators ($a^\dagger a^\dagger a^\dagger a^\dagger, a^\dagger a^\dagger
a^\dagger a, a^\dagger a^\dagger a a, \dots $) vanishes.
\begin{align}
 \delta \langle \phi(q)|
\text{ (normally ordered fourth-order terms)}
|\phi(q)\rangle =0.  \label{eq:d<4th-order terms>=0}
\end{align}

Let us substitute $\hat Q=\hat Q_A+\hat Q_B$ into the 
moving-frame HFB \& QRPA equations in Approach B and 
compare them with those in Approach A.
First, one can see that $\hat Q_B$ does not contribute to the moving-frame HFB equation (\ref{eq:moving-frame HFB}).
\begin{align}
& \delta \langle \phi(q)|\hat H -\lambda \hat N -\partial_q V\hat Q
  |\phi(q)\rangle =0,\notag\\
\Leftrightarrow & \delta \langle \phi(q)|\hat H -\lambda \hat N -\partial_q V\hat Q_A
  |\phi(q)\rangle =0.
\end{align}
As a matter of course, $\hat Q_A$ corresponds to $\hat Q^{(1)}$:
\begin{align}
\hat Q^{(1)} \Leftrightarrow  \hat Q_A.
\end{align}

Next, the moving-frame QRPA equation of $O(p)$ (\ref{eq:moving-frame
QRPA1 }) reads
\begin{align}
 \delta \langle \phi(q)|[\hat H-\lambda \hat N -\partial_q V \hat Q_A-\partial_q V \hat Q_B,
  \hat Q_A+\hat Q_B] 
-\frac{1}{i}B(q)\hat P  |\phi(q)\rangle =0,\notag \\
\Leftrightarrow
 \delta \langle \phi(q)|
[\hat H-\lambda \hat N -\partial_q V \hat Q_A, \hat Q_A]
-\frac{1}{i}B(q)\hat P  
-\partial_q V [\hat Q_B,\hat Q_A]
|\phi(q)\rangle =0.
\label{eq:mfQRPA of O(p) with QB}
\end{align}
Here, 
we have used that $\hat H-\lambda \hat N -\partial_q V \hat Q_A$ does
not contain A-terms,
which follows from  the moving-frame HFB equation~(\ref{eq:moving-frame
HFB}), 
and that the commutator of the normally ordered fourth-order terms of $(a^\dagger, a)$ 
(from the residual interaction part of $\hat H$)
with the B-term does not contribute, i.e.,
\begin{align}
\delta \langle \phi(q)| [\text{normally ordered forth-order terms},
 \text{B-terms}]
|\phi(q)\rangle=0 .
\label{eq:delta<[4th order, B]>=0}
\end{align}
This can easily seen with Eqs. (\ref{eq:[V,B]})--(\ref{eq:[X,B]}),
their Hermitian conjugates, and Eq. (\ref{eq:d<4th-order terms>=0}).

By comparing Eq.(\ref{eq:mfQRPA of O(p) with QB}) with Eq. (\ref{eq:moving-frame QRPA1 with Q3}),
one finds the correspondence as follows.
\begin{align}
\hat Q^{(2)} \Leftrightarrow  i[\hat Q_B,\hat Q_A].
\end{align}

Then, we consider the moving-frame QRPA equation of $O(p^2)$ (\ref{eq:moving-frame QRPA2 }).
From Eq. (\ref{eq:moving-frame QRPA1 }), we have
\begin{align}
 \frac{1}{i}\hat P= \frac{1}{B}[\hat H_M,\hat Q]_A= \frac{1}{B}[\hat H_M,\hat Q_A]_A,
\label{eq: rewritten P}
\end{align}
with
\begin{align}
\hat H_M:=\hat H -\lambda \hat N -\partial_q V \hat Q.
\end{align}
In the second equality in Eq. (\ref{eq: rewritten P}),
we have used  Eqs. (\ref{eq:[B,B]=B}) and (\ref{eq:[V,B]})--(\ref{eq:[X,B]}).

Thus the moving-frame QRPA equation (\ref{eq:moving-frame QRPA2 }) is
rewritten as
\begin{align}
& \delta \langle \phi(q)|
[\hat H -\lambda \hat N -\partial_q V \hat Q_A, \frac{1}{i}\hat P]
-\partial_q V [\hat Q_B, \frac{1}{i}\hat P]
-C(q)\hat Q_A-\partial_q \lambda \hat N \notag\\
&\hspace{9em}
-\frac{1}{2B}
\partial_q V
\left\{[[\hat H-\lambda \hat N -\partial_q V \hat Q,  \hat Q],\hat Q] \right\}
|\phi(q)\rangle =0, \notag \\
\Leftrightarrow &
 \delta \langle \phi(q)|
[\hat H -\lambda \hat N -\partial_q V \hat Q_A, \frac{1}{i}\hat P]
-C(q)\hat Q_A-\partial_q \lambda \hat N \notag\\
&\hspace{6em}
-\frac{1}{B}\partial_q V [\hat Q_B, [\hat H_M,\hat Q_A]_A] 
-\frac{1}{2B}
\partial_q V
[[\hat H_M,  \hat Q],\hat Q] 
|\phi(q)\rangle =0. 
\label{eq:mfQRPA of O(p2) with QB}
\end{align}
The fourth term is rewritten as
\begin{align}
 &\delta \langle \phi(q)|
\frac{1}{B}\partial_q V [[\hat H_M, \hat Q_A]_A, \hat Q_B]|\phi(q)\rangle \notag\\
=&\delta \langle \phi(q)|
\frac{1}{B}\partial_q V [[\hat H_M, \hat Q_A], \hat Q_B]|\phi(q)\rangle,
\end{align}
where we have used that $[\hat H_M, \hat Q_A]$ contains B-terms and
normally ordered fourth-order terms but that they
do not contribute because of Eqs. (\ref{eq:delta<B>=0}) (\ref{eq:[B,B]=B}) and (\ref{eq:delta<[4th order, B]>=0}).
Similarly, one can easily see that
\begin{align}
 &\delta \langle \phi(q)|[[\hat H_M,  \hat Q_B],\hat Q_B] |\phi(q)\rangle =0.
\end{align}

Then, the fourth and fifth terms in Eq. (\ref{eq:mfQRPA of O(p2) with
QB}) are rewritten as
\begin{align}
& \delta \langle \phi(q)|
\frac{1}{B}\partial_q V [[\hat H_M,\hat Q_A],\hat Q_B] 
 -\frac{1}{2B}\partial_q V
[[\hat H_M,  \hat Q_A],\hat Q_B] 
\notag\\
&\hspace{12.5em}
 -\frac{1}{2B}\partial_q V
[[\hat H_M,  \hat Q_B],\hat Q_A]
 -\frac{1}{2B}\partial_q V
[[\hat H_M,  \hat Q_A],\hat Q_A]
|\phi(q)\rangle \notag\\
=&- \delta \langle \phi(q)|
\frac{1}{2B}\partial_q V \left( 
[[ \hat Q_A,\hat Q_B],\hat H_M] 
+[[\hat H_M,  \hat Q_A],\hat Q_A]
\right)
|\phi(q)\rangle 
\notag\\
=& \delta \langle \phi(q)|
-\frac{1}{2B}\partial_q V 
[[\hat H_M,\hat Q_A],\hat Q_A]\notag\\
&+
\frac{1}{2B}\partial_q V 
[\hat H-\lambda\hat N,   [\hat Q_A,\hat Q_B]]
-\frac{1}{2B}(\partial_q V)^2 [[\hat Q_B ,   \hat Q_A],\hat Q_B] |\phi(q)\rangle .
\end{align}
In the second equality, we have used the Jacobi identity.

Finally we obtain
the moving-frame QRPA equation of $O(p^2)$ as follows.
\begin{align}
\delta \langle \phi(q)|&
[\hat H -\lambda \hat N -\partial_q V \hat Q_A, \frac{1}{i}\hat P]
-C(q)\hat Q_A-\partial_q \lambda \hat N \notag\\
&
-\frac{1}{2B}\partial_q V\left(
[[\hat H_M,  \hat Q_A],\hat Q_A] 
-
[\hat H-\lambda\hat N,   [\hat Q_A,\hat Q_B]]\right.\notag\\
&\hspace{8em}\left.+\partial_q V [[\hat Q_B ,   \hat Q_A],\hat Q_B] \right)|\phi(q)\rangle 
=0.
\label{eq:mfQRPA of O(p^2) rewritten} 
\end{align}
It may be noteworthy that, in the derivation of this equation, we have used the
moving-frame QRPA equation of $O(p)$ (\ref{eq:moving-frame QRPA1 }), 
which implies that  
this equation was derived with all the expansions of  $O(1)$,$O(p)$, and
$O(p^2)$ of the equation of collective submanifold (\ref{eq:Eq. of collective submanifold}).

We compare Eq. (\ref{eq:mfQRPA of O(p^2) rewritten}) 
with Eq.(\ref{eq:moving-frame QRPA2 with Q3 }) and find 
\begin{align}
\hat Q^{(3)}-\frac{i}{2} [\hat Q^{(1)}, \hat Q^{(2)}] \Leftrightarrow [[\hat Q_B ,   \hat Q_A],\hat Q_B] .
\end{align}
Because $[\hat Q^{(1)}, \hat Q^{(2)}]$ is a B-term and does not
contribute, we obtain
\begin{align}
\hat Q^{(3)} \Leftrightarrow [[\hat Q_B ,   \hat Q_A],\hat Q_B] =-[\hat Q_B ,   [\hat Q_B,\hat Q_A]] .
\end{align}
Again in this case, one can see the same correspondence as we have seen above
\begin{align}
\hat Q^{(2)} \Leftrightarrow  i[\hat Q_B,\hat Q_A].
\end{align}

\section{Correspondence between higher-order operators and 
the B-part of the first-order operator}
\subsection{The case without pairing correlation}
In the previous section, we have found some correspondence between
the higher-order operators $\hat Q^{(i)}\,(i=2,3)$ in Approach A
and the commutators of the first-order operators $\hat Q_A$ and $\hat
Q_B$ in Approach B.
It implies that it is equivalent to introduce the
B-part of $\hat Q$ to introducing the higher-order operators 
given by this correspondence, at least, at the level of the equations of motion, i.e., the
moving-frame HFB \& QRPA equations.
In this section, we directly derive the correspondence between the
B-part of the first-order operator 
and the higher-order operators by rewriting the state vector
in Approach B.
First, we consider the case where there is no pairing correlation
to illustrate how to derive the relation of $\hat Q_B$ to the
higher-order operators.
The state vectors in the no-pairing case  are  obtained by setting $n=0$
and $\varphi=0$
in Eq. (\ref{eq:state vector}) with $\hat G$ (\ref{eq: G expanded up to 1st}).
The case with pairing correlation is treated in a later subsection.

In Lemma 2 in Ref. \cite{Marumori1980}, it is proven, in the case of no
pairing, that the  unitary operator $e^{i\hat G}$
can be decomposed in the form of
\begin{align}
 e^{i\hat G}=e^{i\hat G_A}e^{i\hat G_B},
\end{align}
where the Hermitian operators $\hat G_A$ and $\hat G_B$ consist of only A-terms and
B-terms, respectively.
In the no-pairing case, the $a^\dagger a^\dagger$($aa$) terms correspond to
the particle-hole pair creation (annihilation) operators, and the $a^\dagger a$ terms
correspond to the particle-scattering and hole-scattering terms.
In Ref. \cite{Marumori1980},
no explicit expressions of $\hat G_A$ and $\hat G_B$ are given.
We shall give explicit expressions for $\hat G = p(\hat Q_A +\hat Q_B)$ below.

In the following, 
we denote $(i\hat Q_A,i\hat Q_B):=(A, B)$ and $(i\hat G_A, i\hat
G_B):=(G_A, G_B)$ to simplify the notation, and
then the state vector is written as
\begin{align}
 e^{p(A + B)}|\phi\rangle =e^{G_A}e^{G_B}|\phi\rangle \label{eq:decomposition}
\end{align}
Here, $G_{A(B)}$ contains only A(B)-terms. 
We shall see below that $G_B$ consists of the $a^\dagger a$ part and a
constant $\theta$. The $a^\dagger a$ part does not contribute to the
state vector because $a^\dagger_\mu a_\nu|\phi(q)\rangle=0$.
Note that the constant term $\theta$ can not be ignored, however.
It depends on $A$ and $B$ as well as $p$, that is, $\theta=\theta(A,B,p)$.
Actually, it is easily shown that, if we omit the constant term 
(and hence $e^{G_B}$)
as below
\begin{align}
e^{p(A+B)}|\phi\rangle=e^{pA+\frac{1}{2}p^2
 A^{(2)}+\frac{1}{6}p^3A^{(3)}+\cdots }|\phi\rangle ,
\end{align}
expand in powers of $p$ and compare the both sides order by order, 
$A^{(3)}$ must contain a B-part.

We first show some formulae (Ref. \cite{Hall2003}) we use below.

\noindent
Baker-Campbell-Hausdorff (BCH) formula:
\begin{align}
 e^{tA}e^{tB}&=\exp \left\{
t(A+B) +\frac{t^2}{2}[A,B] \right.\notag \\
&\left.+\frac{t^3}{12}\left([[A,B],B]+ [A,[A,B]]\right)
+\frac{t^4}{24}[[[B,A],A],B]+\cdots
\right\}.\label{eq:BCH}
\end{align}

\noindent
Zassenhaus formula:
\begin{align}
  e^{t(A+B)}&=e^{tA}e^{tB}e^{-\frac{t^2}{2}[A,B]
 }e^{\frac{t^3}{3!}(2[B,[A,B]]+[A,[A,B]])}\notag\\
&\times e^{-\frac{t^4}{4!}(
 [[A,B],A],A] + 3[[A,B],A],B] + 3[[A,B],B],B])}\cdots.
\label{eq:Zassenhaus}
\end{align}

\noindent
Lie group commutator:
\begin{align}
e^{tA}e^{tB}e^{-tA}&e^{-tB}=e^{\mathcal{U}} \\
\mathcal{U}(t,A,B)&=-\mathcal{U}(t,B,A)\notag \\
&= t^2[A,B]+ \frac{t^3}{2}[(A+B),[A,B]] \notag\\
&+\frac{t^4}{3!}
\left( [[[B,A],A],B]/2+[(A+B), [(A+B), [A,B]]] \right)\cdots,
\label{eq:Lie group commutator 0}
\end{align}
from which we obtain
\begin{align}
 &e^{tA}e^{tB}=e^{tB}e^{tA}e^{\mathcal{U}(-t,A,B)} \notag\\
=&e^{tB}e^{tA}\exp \left\{
t^2[A,B]- \frac{t^3}{2}[(A+B),[A,B]] \right.\notag\\
&\left.+\frac{t^4}{3!}
\left( [[[B,A],A],B]/2+[(A+B), [(A+B), [A,B]]] \right)\cdots \right\}.
\label{eq:Lie group commutator}
\end{align}

The formula (\ref{eq:Lie group commutator 0}) is derived from the BCH formula (\ref{eq:BCH}).
By using these formulae,  
we rewrite $e^{p(A+B)}|\phi\rangle$ to derive the expressions of $G_A$
and $G_B$.
The basic strategy for the derivation is as follows.
\begin{enumerate}
 \item Using the Zassenhaus formula (\ref{eq:Zassenhaus}), transform a sum of Lie algebra
       elements into a product of Lie group
       elements. 
 \item Change the order of a product of Lie group elements 
       using the formula of Lie group commutator (\ref{eq:Lie group
       commutator}). 
       (Shift a B-term to the right
       and an A-term to the left.)
 \item With the BCH formula (\ref{eq:BCH}), transform a product of  Lie group
       elements into a sum of Lie algebra elements. In the sum, there
       appears B-terms, and then go back to (1).
\end{enumerate}
We repeat these steps until we obtain the expression up to the order we need.
Below we take up to $O(p^4)$ and omit the higher-order terms.
We shift $e^{pB}$ to the right as below.
\begin{align}
  &e^{p(A+B)}|\phi \rangle =e^{pA}e^{pB}e^{-\frac{p^2}{2}[A,B]
 }e^{\frac{p^3}{3!}(2[B,[A,B]]+[A,[A,B]])}\notag\\
&\hspace{6em} \times e^{-\frac{p^4}{4!}(
 [[[A,B],A],A] + 3[[[A,B],A],B] + 3[[[A,B],B],B])}\cdots |\phi \rangle
 \notag \\
=&e^{pA}e^{-\frac{p^2}{2}[A,B]}e^{pB}
\exp\left\{ -\frac{p^3}{2} [B, [A,B] ] +\frac{p^4}{4}[B,[B,[A,B]]]
\right\}
e^{\frac{p^3}{6}(\cdots ) } e^{-\frac{p^4}{24}(\cdots)}\cdots|\phi
 \rangle \notag \\
=&e^{pA}e^{-\frac{p^2}{2}[A,B]}
\exp\left\{ -\frac{p^3}{2} [B, [A,B] ] +\frac{p^4}{4}[B,[B,[A,B]]]
\right\}e^{pB}
\exp\left\{-\frac{p^4}{2} [B,[B, [A,B] ]  ]
\right\}
\notag\\
&\times e^{\frac{p^3}{6}(\cdots ) } e^{-\frac{p^4}{24}(\cdots)}\cdots|\phi
 \rangle \notag \\
=&e^{pA}e^{-\frac{p^2}{2}[A,B]}
\exp\left\{ -\frac{p^3}{2} [B, [A,B] ] +\frac{p^4}{4}[B,[B,[A,B]]]
\right\}
\exp\left\{-\frac{p^4}{2} [B,[B, [A,B] ]  ]
\right\}e^{pB}
\notag\\
&\times e^{\frac{p^3}{6}(\cdots ) } e^{-\frac{p^4}{24}(\cdots)}\cdots|\phi
 \rangle \notag \\
=&e^{pA}e^{-\frac{p^2}{2}[A,B]}
\exp\left\{ -\frac{p^3}{2} [B, [A,B] ] -\frac{p^4}{4}[B,[B,[A,B]]]
\right\}
\exp\left\{\frac{p^3}{6}\left(2[B,[A,B]]+[A,[A,B]] \right)\right\}
\notag\\
&\times e^{pB}
\exp\left\{ \frac{p^4}{6}\left(2[B,[B,[A,B]]]+[B,[A,[A,B]]] \right)
\right\}
e^{-\frac{p^4}{24}(\cdots)}\cdots|\phi
 \rangle \notag \\
=&e^{pA}e^{-\frac{p^2}{2}[A,B]}
\exp\left\{
\frac{p^3}{6}\left( [B,[B,A]]+[A,[A,B]] \right)
 -\frac{p^4}{4}[B,[B,[A,B]]]
\right\}\notag\\
&\times 
\exp\left\{ \frac{p^4}{6}\left(2[B,[B,[A,B]]]+[B,[A,[A,B]]] \right)
\right\}
\times e^{-\frac{p^4}{24}(\cdots)}e^{pB}\cdots|\phi
 \rangle \notag \\
%
=&e^{pA}e^{-\frac{p^2}{2}[A,B]}
\exp\left\{
\frac{p^3}{6}\left( [B,[B,A]]+[A,[A,B]] \right) 
+\frac{p^4}{12}[B,[B,[A,B]]]
+\frac{p^4}{6}[B,[A,[A,B]]]
\right\}\notag\\
&\times \exp\left\{ 
-\frac{p^4}{24}\left(
[[[A,B],A],A] + 3[[[A,B],A],B] + 3[[[A,B],B],B]
\right)
\right\}
e^{pB}\cdots|\phi
\rangle .
\end{align}
Noting that
\begin{align}
 [[[A,B],B],B]&=-[[B, [A,B]],B]=[B, [B,[A,B]]], \\
 [[[A,B],A],B]&=-[[A, [A,B]],B]=[B, [A,[A,B]]], \\
 [[[A,B],A],A]&=-[[A, [A,B]],A]=[A, [A,[A,B]]], 
\end{align}
and with the BCH formula~(\ref{eq:BCH}), 
one can rewrite the state vector as
\begin{align}
e^{p(A+B)}|\phi\rangle
& =
\exp\left\{pA
+\frac{p^2}{2}[B,A]
+\frac{p^3}{6}[B,[B,A]]
-\frac{p^3}{12}
[A,[A,B]] 
 \right.\notag\\
&\left.
\hspace{5em}
+\frac{p^4}{24 }[B,[B,[B,A]]]
-\frac{p^4}{24 }[B,[A,[B,A]]]
+\frac{p^4}{12}
[A,[B,[B,A]]]
\right\}
e^{pB}\cdots|\phi
\rangle \notag \\
&=
\exp\left\{pA
+\frac{p^2}{2}[B,A]
+\frac{p^3}{6}[B,[B,A]]
-\frac{p^3}{12}
[A,[A,B]] 
 \right.\notag\\
&\left.
\hspace{6em}
+\frac{p^4}{24 }[B,[B,[B,A]]]
+\frac{p^4}{24 }[B,[A,[B,A]]]
\right\}
e^{pB}\cdots|\phi
\rangle . 
\label{eq:exp(p(A+B))|phi}
\end{align}
We have used that
\begin{align}
 [A,[B,[B,A]]]&=-[B,[[B,A],A]]-[[B,A],[B,A]]=[B,[A,[B,A]]].
\end{align}
The exponent of the first factor contains B-terms in Eq. (\ref{eq:exp(p(A+B))|phi}),
so we decompose it using the Zassenhaus formula (\ref{eq:Zassenhaus}).
\begin{align}
e^{p(A+B)}|\phi\rangle&=
\exp\left\{pA
+\frac{p^2}{2}[B,A]
+\frac{p^3}{6}[B,[B,A]]
+\frac{p^4}{24 }[B,[B,[B,A]]]
 \right.\notag\\
&\left.
\hspace{4em}
-\frac{p^3}{12}[A,[A,B]] 
+\frac{p^4}{24 }[B,[A,[B,A]]]
\right\}
e^{pB}\cdots|\phi
\rangle \notag\\
&=
\exp\left\{pA
+\frac{p^2}{2}[B,A]
+\frac{p^3}{6}[B,[B,A]]
+\frac{p^4}{24 }[B,[B,[B,A]]]
 \right\}\notag\\
&\times \exp\left\{
-\frac{p^3}{12}[A,[A,B]] 
+\frac{p^4}{24 }[B,[A,[B,A]]]
\right\}
\exp\left\{\frac{p^4}{24}[A,[A,[A,B]]] 
]
\right\}
e^{pB}\cdots|\phi
\rangle \notag \\
&=
\exp\left\{pA
+\frac{p^2}{2}[B,A]
+\frac{p^3}{6}[B,[B,A]]
+\frac{p^4}{24 }\left([B,[B,[B,A]]]+[A,[A,[A,B]] \right)
 \right\}\notag\\
&\times \exp\left\{
\frac{p^3}{12}[A,[B, A]]
+\frac{p^4}{24}
[B,[A,[B,A]]]
\right\}e^{pB}|\phi
\rangle +O(p^5)\label{eq:eGAeGB}.
\end{align}
Note that
\begin{align}
 e^{pB}|\phi\rangle = |\phi\rangle,
\end{align}
when $B$ does not contain a constant term and consists of only $a^\dagger a$ terms.
(As mentioned above, it follows from the zeroth-order canonical-variable
condition that $B=i\hat Q_B$ consists of only $a^\dagger a$ terms in the case of the
ASCC method.)

Omitting $e^{pB}$, the state vector reads
\begin{align}
&e^{p(A+B)}|\phi\rangle\notag\\
=&
\exp\left\{pA
+\frac{p^2}{2}[B,A]
+\frac{p^3}{6}[B,[B,A]]
+\frac{p^4}{24 }\left([B,[B,[B,A]]]+[A,[A,[A,B]]]  \right)
 \right\}\notag\\
\times& \exp\left\{
\frac{p^3}{12}[A,[B, A]] 
+\frac{p^4}{24 }
[B,[A,[B,A]]]
\right\}|\phi
\rangle +O(p^5)\label{eq:expG_AexpG_B}.
\end{align}
The exponent of the second exponential factor
is B-terms and corresponds to $G_B$. 
Note that $G_B$ starts from the order of $p^3$. 
The $O(p^3)$ term of $G_A$ does contribute to 
the moving-frame QRPA equation of $O(p^2)$
because the first-order differential operator $\partial_p$ is involved in 
the equation of collective submanifold (\ref{eq:Eq. of collective submanifold}).
On the other hand,
as $\frac{p^3}{12}[A,[B, A]] $ is a B-term,
it does not contribute to the moving-frame QRPA equation of  $O(p^2)$. 
(It contributes to the second-order canonical-variable conditions.)
If there were a term of $O(p^2)$ in $G_B$, it would be involved in
the moving-frame QRPA of $O(p^2)$ in the form of a product with
$ip\hat Q_A$.
However, in the case where the collective coordinate is one-dimensional
and there is no pairing correlation, there can not be $O(p^2)$ terms in
$G_B$ for the following reason. 
What makes B-terms at the second order is a combination of [A-term,A-term] or [B-term,B-term].
The operators we have in this case are $\hat Q_A$ and $\hat Q_B$ only. 
B-terms made of them are $[\hat Q_A,\hat Q_A]$ and $[\hat Q_B,\hat
Q_B]$, and they vanish.
Therefore, it is trivial that there appears no $O(p^2)$ term in $G_B$. 
In the next subsection, 
we shall show that $G_B$ starts from the third order in general,
also in the case with pairing correlation 
and/or the multi-dimensional collective coordinates.

The two factors including B-terms in Eq. (\ref{eq:eGAeGB}) can be
also rewritten as follows.
\begin{align}
&\exp\left\{
\frac{p^3}{12}[A,[B, A]] 
+\frac{p^4}{24 }
[B,[A,[B,A]]]
\right\}e^{pB}\notag\\
=&\exp\left\{
pB+\frac{p^3}{12}[A,[B, A]] 
+\frac{p^4}{24 }
[B,[A,[B,A]]]
+\frac{1}{2}
\left[
\frac{p^3}{12}[A,[B, A]] ,pB
\right]
+ O(p^5)
\right\}\notag\\
=&\exp\left\{
pB+\frac{p^3}{12}[A,[B, A]] 
+\frac{p^4}{24 }
[B,[A,[B,A]]]
+\frac{p^4}{24}
\left[
[A,[B, A]] ,B
\right]
+ O(p^5)
\right\}\notag\\
=&\exp\left\{
pB+\frac{p^3}{12}[A,[B, A]] 
+ O(p^5)
\right\}.
\end{align}
Thus we can rewrite Eq. (\ref{eq:expG_AexpG_B}) as
\begin{align}
 &e^{p(A+B)}|\phi\rangle\notag\\
=&
\exp\left\{pA
+\frac{p^2}{2}[B,A]
+\frac{p^3}{6}[B,[B,A]]
+\frac{p^4}{24 }\left([B,[B,[B,A]]]+[A,[A,[A,B]]]  \right)
 \right\}\notag\\
\times &
\exp\left\{
pB+\frac{p^3}{12}[A,[B, A]]
\right\}|\phi
\rangle +O(p^5).
\end{align}
The fifth- or even higher-order expression can be obtained similarly
with use of the formulae (\ref{eq:BCH})--(\ref{eq:Lie group commutator}).
Also in that case, $G_B$ is written in terms of (multiple) commutators of $A$ and $B$ and can be
rewritten as sum of $a^\dagger a$ and constant terms. 
As $e^{G_B}$ gives just a phase factor, which depends on $p$ as
mentioned above, it does not affect the correspondence between the
higher-order
operators in Approach A and commutators in Approach B listed below.

Finally we find
\begin{align}
 A^{(1)}&=i\hat Q^{(1)}=A\hspace{3.6em}=i\hat Q_A, \\
 A^{(2)}&=i\hat Q^{(2)}=[B,A]\hspace{1.8em}=[i\hat Q_B,i\hat Q_A],   \\
 A^{(3)}&=i\hat Q^{(3)}=[B,[B,A]]=[i\hat Q_B,[i\hat Q_B,i\hat Q_A]],  \\
A^{(4)}&=i\hat Q^{(4)}=[B,[B,[B,A]]]+[A,[A,[A,B]]] \notag\\ 
       &=[i\hat Q_B,[i\hat Q_B,[i\hat Q_B,i\hat Q_A]]]+[i\hat Q_A,[i\hat Q_A,[i\hat Q_A,i\hat Q_B]]] ,   
\end{align}
that is,
\begin{align}
\hat  Q^{(1)}&=\hat Q_A, \label{eq:Q1=QA}\\
\hat  Q^{(2)}&=i[\hat Q_B,\hat Q_A], \label{eq:Q2= QA,QB}\\
\hat  Q^{(3)}&=[\hat Q_B,[\hat Q_A,\hat Q_B]],  \label{eq:Q3= QB,QA,QB} \\
\hat  Q^{(4)}&=-i([\hat Q_B,[\hat Q_B,[\hat Q_B,\hat Q_A]]]+[\hat Q_A,[\hat Q_A,[\hat Q_A,\hat Q_B]]]) \label{eq:Q4}.  
\end{align}
Eqs. (\ref{eq:Q2= QA,QB})--(\ref{eq:Q4}) can be regarded as a 
prescription to determine the higher-order operators.
Once the B-part of $\hat Q$ is given, 
the higher-order operators are determined by the above correspondence.
One prescription to determine $\hat Q_B$ is given in
Ref. \cite{Hinohara2007}.
Note that the inclusion of higher-order operators in $\hat G$
is not equivalent to that of the B-part of the first-order operator.
The B-part of the first-order operator can be always replaced by 
the higher-order A-terms defined by the above correspondence, 
but not vice versa. In general, the higher-order operators can not be expressed by
a single operator $\hat Q_B$.
It is easily understood by counting the numbers of degrees of freedom to
identify $\hat Q_B$ and $\hat Q^{(i)}\,\,(i=1,2,3,\cdots)$.

\subsection{The case with pairing correlation}

The case with pairing correlation can be treated similarly.
Let us denote
$(A^1, A^2):=(i\hat Q_A,i\hat \Theta_A),\,\,\,(B^1, B^2):=(i\hat
Q_B,i\hat \Theta_B)$,
and $(p_1,p_2):=(p,n)$,
and use
the Einstein summation convention to simplify the notation below.
Let us derive the expression up to $O(p_i^3)$.
The state vector in Approach A is given by $e^{i\hat G(q,p,n)} |\phi(q)\rangle$ with
$\hat G(q,p,n)$ expanded to the third order as shown in Eq. (\ref{eq: G expanded up to 3rd}):
\begin{align}
\hat G(q,p,n) &=p \hat Q^{(1)}(q)+n \hat \Theta^{(1)}(q)
+\frac{1}{2}p^2 \hat Q^{(2)}(q) 
+\frac{1}{2}n^2 \hat \Theta^{(2)} (q) +pn \hat X \notag\\
&+\frac{1}{3!}p^3 \hat Q^{(3)}(q) 
+\frac{1}{3!}n^3 \hat \Theta^{(3)}(q) 
+\frac{1}{2}p^2n \hat O^{(2,1)}(q) 
+\frac{1}{2}pn^2 \hat O^{(1,2)}(q) .
\tag{\ref{eq: G expanded up to 3rd}}
\end{align}
We shall transform the state vector in Approach B and compare $\hat G_A$
with $\hat G$ (\ref{eq: G expanded up to 3rd}) as we did in the previous subsection.
Similarly to the previous subsection, we obtain

\begin{align}
&e^{ip(\hat Q_A+\hat Q_B)+in(\hat \Theta_A+\hat\Theta_B) }|\phi(q)\rangle=e^{p_i A^i+
 p_iB^i }|\phi(q)\rangle \notag \\
=& \exp\{p_i A^i + \frac{1}{2}[p_iB^i, p_jA^j] +\frac{1}{6}[p_iB^i,
 [p_jB^j, p_kA^k]] \} \notag \\ 
 &\times
\exp\{-\frac{1}{12}[p_iA^i,[p_jA^j,p_kB^k]]\} e^{p_iB^i}|\phi\rangle
 +O(p^4) \notag\\
=& \exp\{p_i A^i + \frac{1}{2}[p_iB^i, p_jA^j] +\frac{1}{6}[p_iB^i,
 [p_jB^j, p_kA^k]] \} \notag \\ 
 &\times
\exp\{p_iB^i-\frac{1}{12}[p_iA^i,[p_jA^j,p_kB^k]]\}|\phi\rangle
 +O(p^4) \notag\\
=&  \exp\left\{i p \hat Q_A +in\hat \Theta_A 
- \frac{1}{2}p^2[\hat Q_B, \hat Q_A] 
- \frac{1}{2}n^2[\hat \Theta_B, \hat \Theta_A] 
- \frac{1}{2}pn ([\hat \Theta_B, \hat Q_A] +[\hat Q_B, \hat \Theta_A] ) \right.\notag\\
&\hspace{2em} -\frac{i}{6}p^3  [\hat Q_B,  [\hat Q_B, \hat Q_A]] 
 - \frac{i}{6}n^3  [\hat \Theta_B,  [\hat \Theta_B, \hat \Theta_A]]
 \notag \\
&\hspace{2em}- \frac{i}{6}p^2n \left( [\hat Q_B,  [\hat Q_B, \hat \Theta_A]]
+ [\hat \Theta_B,  [\hat Q_B, \hat Q_A]]
+ [\hat Q_B,  [\hat \Theta_B, \hat Q_A]]
 \right) \notag\\
&\left. \hspace{2em} - \frac{i}{6}pn^2 \left( [\hat \Theta_B,  [\hat \Theta_B, \hat Q_A]]
+ [\hat Q_B,  [\hat \Theta_B, \hat \Theta_A]]
+ [\hat \Theta_B,  [\hat Q_B, \hat \Theta_A]]
 \right)
\right\}\notag \\ 
\times &\exp\{-\frac{1}{12}[p_iA^i,[p_jA^j,p_kB^k]]\} e^{i (p\hat Q_B
 +n\hat \Theta_B)}|\phi\rangle
 +O(p^4) \notag 
\end{align}
\begin{align}
=&  \exp\left\{i p \hat Q_A +in\hat \Theta_A 
- \frac{1}{2}p^2[\hat Q_B, \hat Q_A] 
- \frac{1}{2}n^2[\hat \Theta_B, \hat \Theta_A] 
- \frac{1}{2}pn ([\hat \Theta_B, \hat Q_A] +[\hat Q_B, \hat \Theta_A] ) \right.\notag\\
&\hspace{2.em}- \frac{i}{6}p^3  [\hat Q_B,  [\hat Q_B, \hat Q_A]] 
- \frac{i}{6}n^3  [\hat \Theta_B,  [\hat \Theta_B, \hat \Theta_A]]
 \notag \\
&\hspace{2.em}- \frac{i}{6}p^2n \left( [\hat Q_B,  [\hat Q_B, \hat \Theta_A]]
+ [\hat \Theta_B,  [\hat Q_B, \hat Q_A]]
+ [\hat Q_B,  [\hat \Theta_B, \hat Q_A]]
 \right) \notag\\
&\left.\hspace{2.em} - \frac{i}{6}pn^2 \left( [\hat \Theta_B,  [\hat \Theta_B, \hat Q_A]]
+ [\hat Q_B,  [\hat \Theta_B, \hat \Theta_A]]
+ [\hat \Theta_B,  [\hat Q_B, \hat \Theta_A]]
 \right)
\right\}\notag \\ 
 &\times\exp\{i (p\hat Q_B +n\hat \Theta_B)-\frac{1}{12}[p_iA^i,[p_jA^j,p_kB^k]]\} 
|\phi\rangle
 +O(p^4) \label{eq:phi with Q_N and Theta_B}.
\end{align}
One can easily see that the exponents of the first and second factors 
are A-terms and B-terms, and
correspond to  $i\hat G_A$ and to $i\hat G_B$, respectively.
By comparing $\hat G$ (\ref{eq: G expanded up to 3rd}) with $\hat G_A$ in (\ref{eq:phi with Q_N and Theta_B}), 
we read
\begin{align}
&\hat Q^{(2)} = i[\hat Q_B, \hat Q_A], \label{eq:Q2}\\
&\hat \Theta^{(2)} = i[\hat \Theta_B, \hat \Theta_A],  \label{eq:Theta2}\\
&\hat X =\frac{i}{2} \left([\hat Q_B, \hat \Theta_A] +[\hat\Theta_B,\hat Q_A]\right), \\
&\hat Q^{(3)} =[\hat Q_B,[\hat Q_A, \hat Q_B]],\\
&\hat \Theta^{(3)} =[\hat \Theta_B,[\hat \Theta_A, \hat \Theta_B]],\\
&\hat O^{(2,1)} =-\frac{1}{2} 
\left(
[\hat Q_B,[\hat Q_B, \hat\Theta_A]] + [\hat \Theta_B,[\hat Q_B, \hat Q_A]] + [\hat Q_B,[\hat \Theta_B, \hat Q_A]] 
\right), \\
&\hat O^{(1,2)} =-\frac{1}{2} 
\left(
[\hat Q_B,[\hat \Theta_B, \hat\Theta_A]] + [\hat \Theta_B,[\hat Q_B, \hat \Theta_A]] + [\hat \Theta_B,[\hat \Theta_B, \hat Q_A]] 
\right) \label{eq:O12}.
\end{align}
The part of $e^{G_B}|\phi(q)\rangle$ can be rewritten as
\begin{align}
 &\exp\{i (p\hat Q_B +n\hat \Theta_B)-\frac{1}{12} [ p_iA^i, [p_jA^j
 ,p_kB^k ] ]\} 
|\phi\rangle \notag\\
=&
\exp\{-\frac{1}{12}[p_iA^i, [p_jA^j, p_kB^k ] ]\} 
\exp\{i (p\hat Q_B +n\hat \Theta_B)\}
|\phi\rangle
 +O(p^4) \notag\\
=&
\exp\{-\frac{1}{12}[p_iA^i,[p_jA^j,p_kB^k]]\} 
|\phi\rangle
 +O(p^4),
\end{align}
and thus $G_B$ is actually $O(p_i^3)$.
Above we have used that $\hat \Theta_B$ does not contain a constant
term,
which follows from the zeroth-order canonical-variable condition,
\begin{align}
 \langle \phi(q)|\hat \Theta (q)| \phi(q)\rangle 
=\langle \phi(q)|\hat \Theta_B (q)| \phi(q)\rangle
=0.
\end{align}
Here we have considered the case where the collective coordinate is one-dimensional.
One can easily see that $G_B$ is $O(p^3)$ also in the multi-dimensional case.
Hence, $G_B$ does not contribute to the moving-frame HFB \& QRPA 
equations up to $O(p^2)$.

\section{Inertial mass}
As shown in Ref. \cite{Sato2017}, when the higher-order operators are
included, 
the collective Hamiltonian is given by
\begin{align}
\mathcal{H}(q,p,n) &=  V(q)+\frac{1}{2}B(q)p^2 +\lambda n +\frac{1}{2}D(q)n^2, \\
 V(q)&=\langle\phi (q)|\hat H |\phi(q)\rangle, \\
 B(q)&=\langle\phi (q)|[\hat H, i\hat  Q^{(2)}] |\phi(q)\rangle 
      -\langle\phi (q)|[[\hat H, \hat  Q^{(1)}],\hat Q^{(1)}]
 |\phi(q)\rangle, 
\label{eq:B(q)}\\
 \lambda(q)&=\langle\phi (q)|[\hat H,i\hat \Theta^{(1)}] |\phi(q)\rangle, \\
 D(q)&=\langle\phi (q)|[\hat H, i\hat  \Theta^{(2)}] |\phi(q)\rangle 
      -\langle\phi (q)|[[\hat H, \hat  \Theta^{(1)}],\hat \Theta^{(1)}] |\phi(q)\rangle .
\end{align}
The second-order operators $\hat Q^{(2)}$and $\hat \Theta^{(2)}(q)$
contribute to the inertial functions $B(q)$ and $ D(q)$, respectively.

We compare the inertial mass $B(q)$ in the two approaches:
one is Approach A with the second-order operators defined by
(\ref{eq:Q2}),
 and the other Approach B.
We confirm that the inertial masses obtained in
the two approaches coincide with each other.
Substituting Eq. (\ref{eq:Q2}) into Eq. (\ref{eq:B(q)}), we obtain the inertial mass $B(q)$ 
as
\begin{align} 
 B(q)&=-\langle\phi (q)|[\hat H, [\hat Q_B, \hat Q_A]] |\phi(q)\rangle 
      -\langle\phi (q)|[[\hat H, \hat  Q_A],\hat Q_A] |\phi(q)\rangle \notag\\
&=\langle\phi (q)|[\hat Q_B, [\hat Q_A, \hat H]] |\phi(q)\rangle 
  +\langle\phi (q)|[\hat Q_A, [\hat H, \hat Q_B]] |\phi(q)\rangle \notag\\
&\hspace{13em}      -\langle\phi (q)|[[\hat H, \hat  Q_A],\hat Q_A] |\phi(q)\rangle \notag\\
&=
  -\langle\phi (q)|[[\hat H, \hat Q_B],\hat Q_A] |\phi(q)\rangle 
      -\langle\phi (q)|[[\hat H, \hat  Q_A],\hat Q_A] |\phi(q)\rangle
 \notag\\
&\hspace{13em }+\langle\phi (q)|[[\hat H,\hat Q_A],\hat Q_B] |\phi(q)\rangle, 
\end{align}
where we denoted $\hat Q^{(1)}$ by $\hat Q_A$ and used the Jacobi identity.

Next, in Approach B, the inertial mass $B(q)$ is given by
\begin{align} 
 B(q)&=-\langle\phi (q)|[[\hat H, \hat  Q_A+\hat Q_B],\hat Q_A+\hat Q_B] |\phi(q)\rangle \notag\\
&=-\langle\phi (q)|[[\hat H, \hat  Q_A] ,\hat Q_A] |\phi(q)\rangle 
-\langle\phi (q)|[[\hat H,   \hat Q_B],\hat Q_A] |\phi(q)\rangle \notag\\
&\ \ \ \, -\langle\phi (q)|[[\hat H, \hat  Q_A], \hat Q_B] |\phi(q)\rangle 
-\langle\phi (q)|[[\hat H, \hat  Q_B], \hat Q_B] |\phi(q)\rangle
 \label{eq:hinohara B}
\end{align}
Noting that
 $\hat Q_B|\phi\rangle =0$,
we find that both of the two approaches give the same result,
\begin{align} 
 B(q)&=-\langle\phi (q)|[[\hat H, \hat  Q_A] ,\hat Q_A] |\phi(q)\rangle 
-\langle\phi (q)|[[\hat H,   \hat Q_B],\hat Q_A] |\phi(q)\rangle .
\label{eq:B(q)}
\end{align}
One can easily see that such is the case with the inertial function $D(q)$.
The second term is the contribution from the second-order collective
coordinate operator $\hat Q^{(2)}$ in Approach A and that from the B-part of the first-order
operator $\hat Q_B$ in Approach B.

When the ASCC method without pairing correlation 
is applied to the translational motion, 
the second term in Eq. (\ref{eq:B(q)}) vanishes for the following reason.
For the translational motion, $\partial_q V(q)=0$ and the moving-frame
Hamiltonian $\hat H_M$ reduces to the Hamiltonian $\hat H$.
Then, $\hat H_M=\hat H$ does not contain A-terms in the quasiparticle
representation, and the expectation value $\langle\phi (q)|[[\hat H,
\hat Q_B],\hat Q_A] |\phi(q)\rangle$ vanishes because 
$[\hat H,\hat Q_B]$ consists of B-terms and normally ordered forth-order terms.
This can be understood as the reason why the correct mass was obtained without including the B-part
of $\hat Q$ in Ref. \cite{Wen2016}.
%

\section{Concluding remarks}
In this paper, 
we studied the role of $a^\dagger a$ terms in $\hat G$ of the state vector $e^{i\hat G}|\phi\rangle$
in the context of the ASCC method.
We have shown that 
the B-part of the first-order collective coordinate operator 
can be rewritten as
the higher-order operators consisting of only
A-terms in the adiabatic expansion of $\hat G$.
We have given the explicit expressions of the corresponding higher-order
operators, which are written
in terms of (multiple) commutators of the A-part and B-part of the
first-order collective operators.
Once the B-part of the first-order operator is given, the corresponding
higher-order operators are automatically determined.
Thus, this correspondence serves as a prescription to determine the
higher-order collective operators in the adiabatic expansion.

As mentioned above, it is not equivalent including the B-part at the
first order to including the higher-order operators of the adiabatic expansion.
As shown in Ref. \cite{Sato2017}, the gauge symmetry in the ASCC method
is (partially) broken by two sources, i.e., the decomposition of the
equation of motion depending on the order of $p$ 
and the truncation of the adiabatic expansion.
The gauge symmetry broken by the truncation can be conserved by including
higher-order operators up to sufficiently high order.
However, in the case where only the first-order operators are taken into
account, the gauge symmetry of the canonical-variable conditions is broken,
even if the B-part of the first-order operator is introduced.

As discussed  in Ref. \cite{Sato2017} and this paper,
the higher-order operators contribute to the moving-frame QRPA equations
and the inertial function, and thus they may affect the low-lying states
physically.
So far, we have considered the moving-frame equations  up to $O(p^2)$
in the formulation of the ASCC method.
When the higher-order operators are taken into account,
one may need to solve the moving-frame equation(s) of $O(p^i)\,\,(i\geq 3)$
to determine them.
However, in general, it may not be easy to solve such higher-order
moving-frame equations self-consistently, and one may need an
alternative way to determine the higher-order operators without solving the higher-order
moving-frame equations.
The correspondence we have seen in this paper gives one prescription.
With this prescription, what one has to do is to give the B-part of
the first-order collective operator.
One prescription to determine the B-part is already 
given by Hinohara et al (Refs. \cite{Hinohara2007, Hinohara2008, Hinohara2009}), and the B-part of the
collective coordinate operator $\hat Q_B(q)$ is determined
by requiring $[\hat Q,\hat N]=0$ for the gauge symmetry of the
moving-frame HFB \& QRPA equations to be conserved.
(In this case, the gauge symmetry of the canonical-variable conditions
is not completely conserved. This should be regarded as an approximate
way to conserve the gauge symmetry.)


In the case without pairing correlation,
there exists no gauge symmetry (Ref. \cite{Sato2015}).
Therefore, one does not need to introduce the B-part 
in order to retain the gauge symmetry.
However, with the B-part included, 
one can take into account in an effective way the contribution from the higher-order
operators,
which do contribute to the moving-frame QRPA equations and inertial
mass.
When there is no pairing correlation, $[\hat Q, \hat N]=0$ and
a prescription other than that by Hinohara et al is necessary.
It would be interesting to investigate other possible prescriptions and
how meaningful the contribution from the higher-order operators is.
A possible prescription will be studied in a future publication.
As discussed in Sect. 4,
for the translational motion without pairing correlation,
the correct mass can be reproduced with neither 
B-terms nor higher-order terms (Ref. \cite{Wen2016}).
However, for large-amplitude collective motion in general,
the higher-order terms or B-terms contribute to the inertial mass
and the moving-frame (Q)RPA equations.

\section*{Acknowledgment}
The author thanks K. Matsuyanagi, T. Nakatsukasa, N. Hinohara, and
M. Matsuo for fruitful discussions and comments.

\appendix

\section{Some commutators of fermion operators}
Here we show some formulae of commutators involving the second- and
forth-order fermion operators, which are useful for understanding of the derivation in
this paper.

\begin{align}
 [a_\alpha a_\beta, a_\gamma^\dagger a_\delta^\dagger]&= 
\delta_{\alpha\delta}\delta_{\beta\gamma} -\delta_{\alpha\gamma}\delta_{\beta\delta}
-\delta_{\beta\gamma}a_\delta^\dagger a_\alpha +\delta_{\beta\delta}a_\gamma^\dagger
a_\alpha +\delta_{\alpha\gamma} a_\delta^\dagger a_\beta
 -\delta_{\alpha\delta} a_\gamma^\dagger a_\beta, \label{eq:[aa,a+a+]}\\
 [a^\dagger_\alpha a_\beta, a_\gamma^\dagger
 a_\delta^\dagger]&=\delta_{\beta\gamma}a_\alpha^\dagger a_\delta^\dagger - \delta_{\beta\delta}a_\alpha^\dagger a_\gamma^\dagger  ,\\
 [a^\dagger_\alpha a_\beta, a_\delta a_\gamma]&=
 \delta_{\alpha\gamma}a_\beta a_\delta - \delta_{\alpha \delta}a_\beta a_\gamma         ,\\
 [a^\dagger_\alpha a_\beta, a_\gamma^\dagger a_\delta]&=
 \delta_{\beta\gamma} a^\dagger_\alpha a_\delta -\delta_{\alpha \delta}
 a_\gamma^\dagger a_\beta . \label{eq:[a+a,a+a]}
\end{align}

\begin{align}
 [a^\dagger_\alpha a_\beta^\dagger a_\delta^\dagger  a_\gamma^\dagger, a_\mu^\dagger a_\nu]
&=
-\delta_{\nu\alpha} a^\dagger_\mu a_\beta^\dagger  a_\delta^\dagger a_\gamma^\dagger  
+\delta_{\nu\beta}  a^\dagger_\mu a_\alpha^\dagger a_\delta^\dagger a_\gamma^\dagger
-\delta_{\nu\delta} a^\dagger_\mu a_\alpha^\dagger a_\beta^\dagger  a_\gamma^\dagger    
+\delta_{\nu\gamma} a^\dagger_\mu a_\alpha^\dagger a_\beta^\dagger
 a_\delta^\dagger ,
\label{eq:[V,B]}
   \\
 [a^\dagger_\alpha a_\beta^\dagger a_\delta^\dagger  a_\gamma, a_\mu^\dagger a_\nu]
&=
\delta_{\mu\gamma}  a^\dagger_\alpha a_\beta^\dagger  a_\delta^\dagger a_\nu  
-\delta_{\nu\alpha} a^\dagger_\mu    a_\beta^\dagger  a_\delta^\dagger a_\gamma
+\delta_{\nu\beta}  a^\dagger_\mu    a_\alpha^\dagger a_\delta^\dagger a_\gamma
-\delta_{\nu\delta} a^\dagger_\mu    a_\alpha^\dagger a_\beta^\dagger
 a_\gamma ,\\
%
%
 [a^\dagger_\alpha a_\beta^\dagger a_\delta  a_\gamma, a_\mu^\dagger a_\nu]
&=
 \delta_{\mu\gamma} a^\dagger_\alpha a_\beta^\dagger  a_\delta a_\nu 
-\delta_{\mu\delta} a^\dagger_\alpha a_\beta^\dagger  a_\gamma  a_\nu
-\delta_{\nu\alpha} a^\dagger_\mu    a_\beta^\dagger  a_\delta a_\gamma
+\delta_{\nu\beta}  a^\dagger_\mu    a_\alpha^\dagger  a_\delta a_\gamma .   
\label{eq:[X,B]}
\end{align}

From Eqs. (\ref{eq:[V,B]})--(\ref{eq:[X,B]}) and their Hermitian
conjugates, it is easily seen that the commutators between
normally ordered forth-order terms with $a^\dagger a$ terms give 
normally ordered forth-order terms.

%

%


\begin{thebibliography}{10}
\bibitem{Thouless1960}
D. J. Thouless, Nucl. Phys.  {\bf 21}, 225 (1960).

\bibitem{Marumori1980}
T.~Marumori, T.~Maskawa, F.~Sakata, and A.~Kuriyama, Prog. Theor. Phys. {\bf 64}, 1294 (1980).

\bibitem{Rowe1980}
D.~J.~Rowe, A.~Rymann, G.~Rosensteel, Phys. Rev. A {\bf 22}, 2362 (1980).

\bibitem{Ring1977}
P. Ring, P. Schuck, Nucl. Phys. A {\bf 292}, 20 (1977).

\bibitem{Suzuki1983}
T.~Suzuki, Nucl. Phys. A {\bf 398}, 557 (1983).


\bibitem{Sato2017}
K. Sato, Phys. Theor. Exp. Phys. {\bf 2017}, 033D01 (2017).

\bibitem{Matsuo2000}
M. Matsuo, T. Nakatsukasa, and K. Matsuyanagi, Prog. Theor. Phys. {\bf 103}, 959 (2000).


\bibitem{Ring1980}
P. Ring and P. Schuck, 
\newblock {\em {The Nuclear Many-body Problems}},
\newblock  (Springer-Verlag Berlin Heidelberg, 1980).


\bibitem{Hinohara2007}
N. Hinohara, T. Nakatsukasa, M. Matsuo, and K. Matsuyanagi,
  Prog. Theor. Phys. , {\bf 117}, 27 (2007).

\bibitem{Hinohara2008}
N. Hinohara, T. Nakatsukasa, M. Matsuo, and K. Matsuyanagi,
  Prog. Theor. Phys. , {\bf 119}, 59 (2008).

\bibitem{Hinohara2009}
N. Hinohara, T. Nakatsukasa, M. Matsuo, and K. Matsuyanagi,
  Phys. Rev. C {\bf 80}, 014305 (2009).

\bibitem{Matsuyanagi2010}
K. Matsuyanagi, M. Matsuo, T. Nakatsukasa, N. Hinohara, and K. Sato, J. Phys. G {\bf 37},  064018 (2010).

\bibitem{Nakatsukasa2016}
T. Nakatsukasa, K. Matsuyanagi, M. Matsuo, K. Yabana, Rev. Mod. Phys.  {\bf 88},  045004 (2016).

\bibitem{Klein1991}
A.~Klein, N.~R.~ Walet, and G~{Do Dang}, Ann. Phys. {\bf 208},
  90 (1991).

\bibitem{Hinohara2010b}
N. Hinohara, K. Sato, T. Nakatsukasa, M. Matsuo, and K. Matsuyanagi, Phys. Rev. C {\bf 82}, 064313 (2010).

\bibitem{Sato2011}
K. Sato and N. Hinohara, Nucl. Phys. A {\bf 849}, 53 (2011).

\bibitem{Watanabe2011}
H.~Watanabe et al. Phys. Lett. B {\bf 704}, 270 (2011).

\bibitem{Hinohara2011a}
N. Hinohara and Y. Kanada-En'yo, Phys. Rev. C {\bf 83},
  014321 (2011).

\bibitem{Hinohara2011}
N. Hinohara, K. Sato, K. Yoshida, T. Nakatsukasa, M.
  Matsuo, and K. Matsuyanagi, Phys. Rev. C {\bf 84}, 061302
  (2011).

\bibitem{Hinohara2012}
N.~Hinohara, Z.~P. Li, T.~Nakatsukasa, T.~Nik{\v{s}}i{\'{c}}, and D.~Vretenar,
Phys. Rev. C {\bf 85}, 024323 (2012).

\bibitem{Yoshida2011}
K. Yoshida and N. Hinohara, Phys. Rev. C {\bf 83}, 1 (2011).

\bibitem{Sato2012}
K. Sato, N. Hinohara, K. Yoshida, T. Nakatsukasa, M.
  Matsuo, and K. Matsuyanagi, Phys. Rev. C {\bf 86}, 24316
  (2012).

\bibitem{Dirac1950}
P.~A.~M. Dirac, Can. J. Math. {\bf 2}, 129 (1950).

\bibitem{Anderson1951}
J.~L. Anderson and P.~G. Bergmann, Phys. Rev. {\bf 83},
  1018 (1951).

\bibitem{Dirac1964}
P.~A.~M. Dirac,
\newblock {\em {Lectures on Quantum Mechanics}},
\newblock  (Belfer Graduate School of Science Monographs Series, Yeshiva
  University, New York, 1964).

\bibitem{Sato2015}
K. Sato, Prog. Theor. Exp. Phys. {\bf 2015}, 123D01 (2015).


\bibitem{Wen2016}
K. Wen, T. Nakatsukasa, Phys. Rev. C {\bf 94}, 054618 (2016).


\bibitem{Villars1977}
F. Villars, Nucl. Phys. A {\bf 285}, 269 (1977).

\bibitem{Baranger1978}
M. Baranger, M. V{\'e}n{\'e}roni, Ann. Phys. {\bf 114}, 123 (1978).


\bibitem{Hall2003}
B. C. ~Hall,
\newblock {\em {Lie Groups, Lie Algebras, and Representations}},
\newblock  (Springer New York, Second Edition, 2015).


\end{thebibliography}
\end{document}